\documentclass[runningheads]{llncs}
\usepackage[T1]{fontenc}
\usepackage{graphicx}
\usepackage[dvipsnames]{xcolor}
\usepackage{mathtools}
\DeclarePairedDelimiter\abs{\lvert}{\rvert}
\begin{document}
\title{Structure and Context of Retweet Coordination in the 2022 U.S. Midterm Elections}
\titlerunning{Structure of Coordination}
%
\author{David Axelrod \orcidID{0000-0002-9840-8711} (\email{daaxelro@iu.edu}) \and \\
John Paolillo\ \orcidID{0009-0000-0876-5778} (\email{paolillo@indiana.edu})}
\authorrunning{D. Axelrod, et al.}

\institute{Indiana University Observatory on Social Media}

\maketitle              
\begin{abstract}
The ability to detect coordinated activity in communication networks is an ongoing challenge. Prior approaches emphasize considering any activity exceeding a specific threshold of similarity to be coordinated. However, identifying such a threshold is often arbitrary and can be difficult to distinguish from grassroots organized behavior. In this paper, we investigate a set of Twitter retweeting data collected around the 2022 US midterm elections, using a latent sharing-space model, in which we identify the main components of an association network, thresholded with a $k$-nearest neighbor criterion. This approach identifies a distribution of association values with different roles in the network at different ranges, where the shape of the distribution suggests a natural place to threshold for coordinated user candidates. We find coordination candidates belonging to two broad categories, one involving music awards and promotion of Korean pop or Taylor Swift, the other being users engaged in political mobilization. In addition, the latent space suggests common motivations for different coordinated groups otherwise fragmented by using an appropriately high threshold criterion for coordination.
\end{abstract}

\section{Introduction}

A growing body of research has been motivated to understand the potential for manipulating users on social media platforms through coordinated efforts known as influence campaigns \cite{tuckerRussia,pierriUkr,zannettou2019disinformation}. While much of this research benefited from ground truth data provided from social media platforms following congressional hearings on the 2016 elections \cite{nwala2023language,cima2024coordinated,saeed2024unraveling}, the ability to identify other campaigns operating on social media platforms remains a prerequisite to studying other cases of influence campaigns.

Previous work to identify coordinated users has provided evidence that high levels of similarity between users in their social media behaviors can function as an effective signal of coordination. In much of this work, there is a binary classification for coordinated and non-coordinated users by choosing a high cosine similarity or Jaccard coefficient as a threshold beyond which users are considered coordinated \cite{Pacheco_2020,Pacheco_2021,nizzoli2021coordinated,magelinski2022synchronized}. This approach has been demonstrated with respect to the promotion of influencers, hashtags, websites, and account handle swapping, to name some of the most prominent behavior dimensions considered \cite{Pacheco_2021,coURL,magelinski2022synchronized,tardelli2023temporal}. 

However, there are still open questions on how to approach identification and characterization of campaigns. One of these questions is how to achieve meaningful separation between truly coordinated users and organic users that may look suspicious due to a natural tendency for homophily in social networks. Because motivated users may consolidate around common objectives as an emergent phenomenon, it is often difficult to discern these users from coordinated users that are connected by some latent organizational structure or purpose. The crux of the problem is that there are two different processes, one bottom-up the other top-down, which both produce high similarity among users. Applying an extremely high similarity threshold may succeed in isolating some proportion of coordinated users, but this approach has been criticized for the potential to classify coordinated users as organic ones \cite{mutlifacet2020,nizzoli2021coordinated}. On the other hand, alternative approaches that are meant to extract a network backbone for networks derived from complex, multi-scale systems do not satisfactorily align with the task of isolating coordinated users. These methods have the potential to retain low-weighted edges between users when these edges satisfy a given pruning criterion. Another open question, less critical but still very important for our comprehension of campaigns, is achieving a clear understanding of coordination campaign objectives. 

This paper starts to address these challenges through a case study of retweeting behavior in the 2022 United States midterm elections. Our analysis begins by constructing a latent space model of sharing behavior in the data. This builds on work that utilizes follow or interaction networks to infer latent ideological relationships between users \cite{conover2011predicting,barbera2015birds,Barbera_2,wong2016quantifying}. In particular, we expand on studies utilizing Singular Value Decomposition (SVD) to model users in terms of their retweet behavior \cite{PolPolar2023}, allowing us to analyze the main retweeting trends among organic users. Next, we approach the coordination classification problem as a variable association inference problem by applying a measure of statistical association: Cramer's Phi, symbolized and hereafter referred to as $\phi$. This measure is related to the Pearson product-moment correlation and to Pearson's chi-squared test for independence \cite{bergsma2013bias}, and so has well-understood properties for inferring relationships between users. Finally, we engage with the output from both analyses for a broad understanding of the context in which the coordinated activity took place. 

\section{Methods}
\subsection{Data}
This analysis employs data from a published Twitter dataset collected during the 2022 United States midterm elections \cite{aiyappa2023multi}. The query for this data was established using a snowball sample method. Data collection began using a limited set of topic-relevant terms e.g., ``midterms'', ``'elections', etc.; the initial collection was then examined for additional relevant terms to expand the search query; for more details about this process see \cite{aiyappa2023multi}. We limit our analysis to retweet data collected during the most active week this query was run, from November 5 to November 11. Because we are interested in analyzing motivated users and content distributed by organic movements or campaigns, we consider users who retweet twenty or more tweets during this period and tweets that are shared by at least ten unique users. This results in 1,663,946 retweets from 73,097 unique users. 

\subsection{Neighborhood Association} 
An association matrix was constructed by first identifying each node's three nearest neighbors, determined by the cosine similarity for user pairs. The immediate neighbors of each node are sufficient for detecting coordination, because having even one very high association with a neighbor is both necessary and sufficient for coordinated activity. Furthermore, such nearest-neighbor association is likely to be transitive within clusters of coordinating users. Remaining entries of the association matrix are set to zero. 

The association $\phi$ is computed for the edges in the nearest neighbor graph by considering each connected pair of users, $u_1$ and $u_2$, and counting retweets for each of the four cells of the contingency table defined in Table \ref{tab:2by2}; $\phi$ is computed in as in (\ref{eq:phiassoc}) from the counts $a$, $b$, $c$, and $d$; $\phi$ may also be computed equivalently from the chi-squared test statistic \cite{bergsma2013bias}. 

\begin{table}[]
    \centering
    \begin{tabular}{c|cc}
         & $u_2$ yes &  $u_2$ no \\
        \hline
        $u_1$ yes &  $a$ &  $b$ \\
        $u_1$ no &  $c$ & $d$
    \end{tabular}
    \caption{Two-by-two table for comparison of retweets by user pairs.}
    \label{tab:2by2}
\end{table}

\begin{equation} \label{eq:phiassoc}
    \phi = \abs*{ \frac{a \cdot d - b \cdot c}{\sqrt{a \cdot b \cdot  c \cdot d}} }
\end{equation}

\subsection{Latent Sharing Space} 
We construct the latent sharing space through SVD of the entire retweeter-tweet matrix. After binarizing all count values, we perform double centering by subtracting the expected values for each cell under a model of independence for users and tweets, as in (\ref{eq:delta}); the deviances are the corresponding centered values, which we arranged in a user-by-tweet matrix. 

\begin{equation} \label{eq:delta}
    \delta_{user,tweet} = obs_{user,tweet} - \frac{ n_{user} \cdot n_{tweet}}{n_{ total}} 
\end{equation}

We then compute the SVD of the centered matrix to identify a latent retweet sharing space, from which we extract dimension scores for the users. The distribution of users within this space corresponds to correlated retweeting activity among the users. Users were then clustered using density-based clustering with HDBSCAN\cite{mcinnes2017accelerated} after applying row-wise L2-normalization. The resulting four clusters have 9,652 (A), 20,965 (B), 12,764 (C), and 29,278 (D) members, respectively. Only 438 users were identified as outliers with respect to these clusters.

\section{Results}

\subsection{Association}
\begin{figure}
    \centering
    \includegraphics[width=1\textwidth]{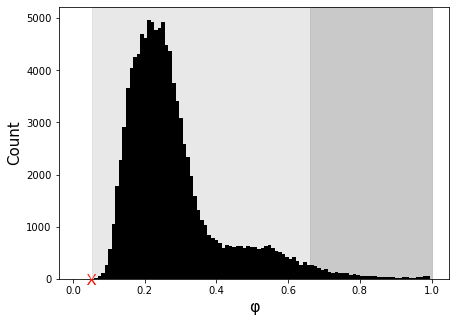}
    \caption{Frequencies of edge weights between pairs of users for which $\phi$ was measured. Yellow region corresponds to values of $\phi$ the we treat as an indicator of coordination. Black and yellow region correspond to critical $\phi$ values derived for a 0.0001 significance level and Bonferroni corrected by the number of user comparisons made (109,118). Using statistical significance as a threshold mechanism would incorrectly suggest almost all our users are coordinated. Instead, we utilize the bimodality of the distribution as seen here and in Figure \ref{fig:phi_thresh}.}
    \label{fig:phi_pv}
\end{figure}

If $\phi$ were computed for all user pairs, the resulting distribution would be heavily skewed with a large number of values near zero, providing no clear guide to thresholding for identifying coordination.  In contrast, non-zero values of $\phi$  for the 3-nearest neighbor pairs are distributed as shown in Figure \ref{fig:phi_pv}.  Because the computed associations are conditioned on pre-determined similarity of the users,  it is inappropriate to use reference values of $\phi$ assuming independence, as would be done in significance testing. In any event, using an arbitrarily strict threshold fails to reject most values from consideration, as indicated by the location of $p_{crit}$ indicated by the red {\color{red} x} on the $x$-axis of Figure \ref{fig:phi_pv}. However, the distribution is clearly bi-modal, with a large mode of weaker associations and a smaller mode of stronger ones. Hence, users appear to have two distinct kinds of network-structuring retweeting activities. Since we expect coordination to require close associations, we consider the upper range worth investigating for coordination candidates.

\begin{figure}
    \centering
    \includegraphics[width=1\textwidth]{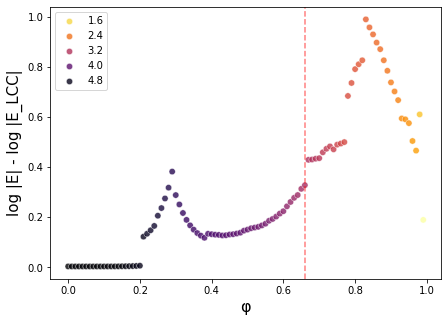}
    \caption{Log ratio of edges in the network to edges in the largest connected component for increasing threshold of $\phi$ ($x$-axis). As the threshold is increased, edges falling below the threshold are removed. Two peaks are observed with increasing trends corresponding to threshold values where the largest connected component aggressively decomposes into smaller components. Point colors encode the edge counts on log scale, as indicated in the key.}
    \label{fig:phi_thresh}
\end{figure}

\begin{figure*}
    \centering
    \includegraphics[width=1\textwidth]{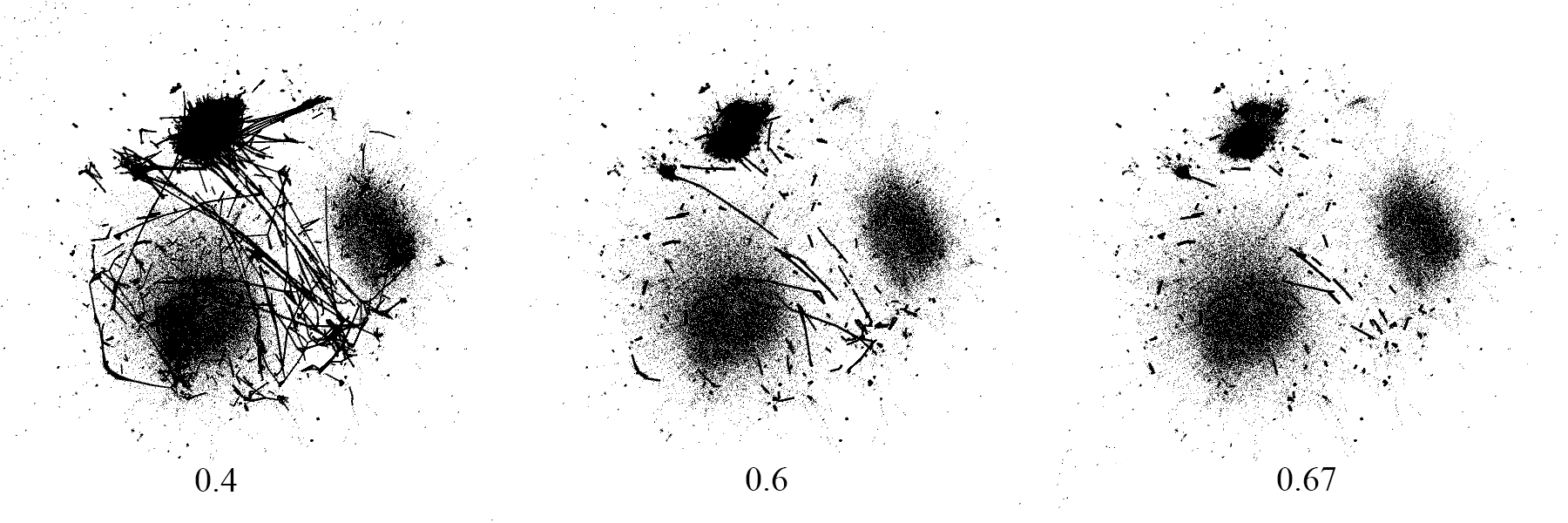}
    \caption{Association network thresholded at three $\phi$ values. Our chosen threshold is represented by the last plot, $\phi \geq 0.67$.   }
    \label{fig:edge_thresh}
\end{figure*}

To provide insight regarding the roles of different levels of association to network structure, we consider the effect of thresholding network ties at different association levels on the connectedness of the resulting network, for 100 possible threshold values between 0 and 1. These are plotted on the $x$ axis of Figure \ref{fig:phi_thresh}, against the the log ratio of total edges in the graph to edges in the largest connected component, on the $y$ axis. The bimodality recurs, but shifted rightward somewhat. In addition, the smaller mode of Figure \ref{fig:phi_pv} is more highly connected, reaching a peak near 1 (i.e., a 10:1 ratio) just below an association of 0.8. There are also three discontinuities: one between 0.7 and 0.6, one between 0.75 and 0.8, and one just above 0.8. These breaks suggest distinct network components connected with different association strengths, that dissolve when the association threshold is raised.  

\begin{figure}
    \centering
    \includegraphics[width=1\textwidth]{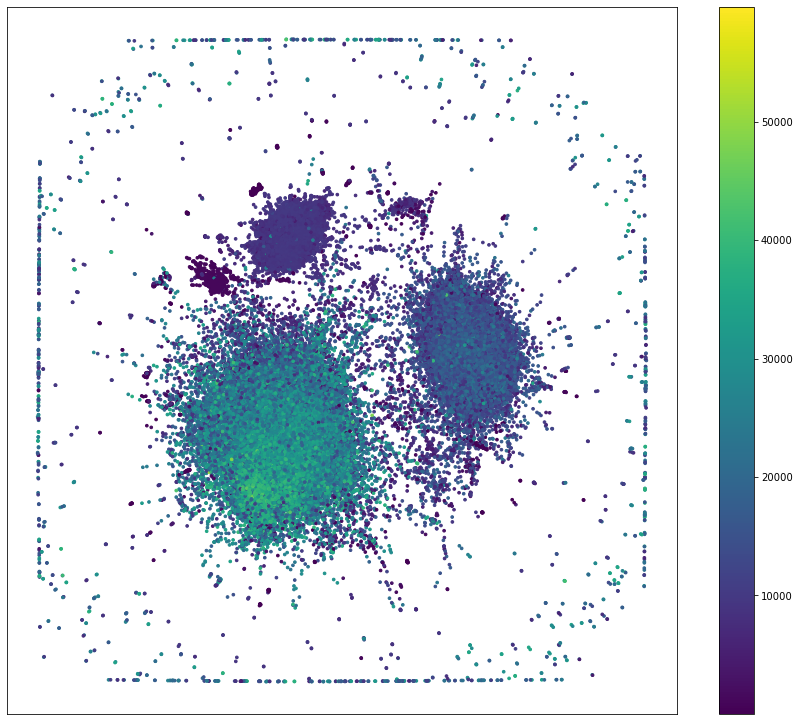}
    \caption{K-nearest neighbor network (k=3) where edges are weighted by $\phi$ and nodes are colored according to their degree in the retweeter-retweeter unimodal projection of the original retweeter-tweet matrix.}
    \label{fig:phi_netd}
\end{figure}

This, also is confirmed in Figure \ref{fig:edge_thresh} in which the 3-nearest neighbor network was first plotted using the Force Atlas 2 layout of Gephi \cite{ICWSM09154}, before thresholding at various levels of $\phi$. Among a scatter of smaller clusters, there are three larger clusters in the network, which we describe further in the latent space analysis below; setting  $\phi \geq 0.4$ leaves most of the network intact, while at $\phi \geq 0.6$ to $\phi \geq 0.67$, bridges between and within each of the main clusters start to disappear, corresponding to our assumption that coordination will take place within clusters, rather than as bridges between them. Figure \ref{fig:phi_net}, employs the same layout, but with node color highlighting each node's strongest association value. Nodes with high association scores gravitate toward the centers of these three clusters, although the smallest of the three, toward the top of the layout, has higher overall association than in the other two clusters. This contrasts with the distribution of node degree, displayed in Figure \ref{fig:phi_netd}, in which a user's degree is dominated by the size of the cluster they belong to -- those of the largest cluster tend to have higher degree, while those in smaller cluster have lower degree. Hence, the three clusters are characterized by shared retweeting behaviors that lead to different structural properties among the three clusters.

\subsection{Decomposition to latent space using SVD}
\begin{figure*}
    \centering
    \includegraphics[width=1\textwidth]{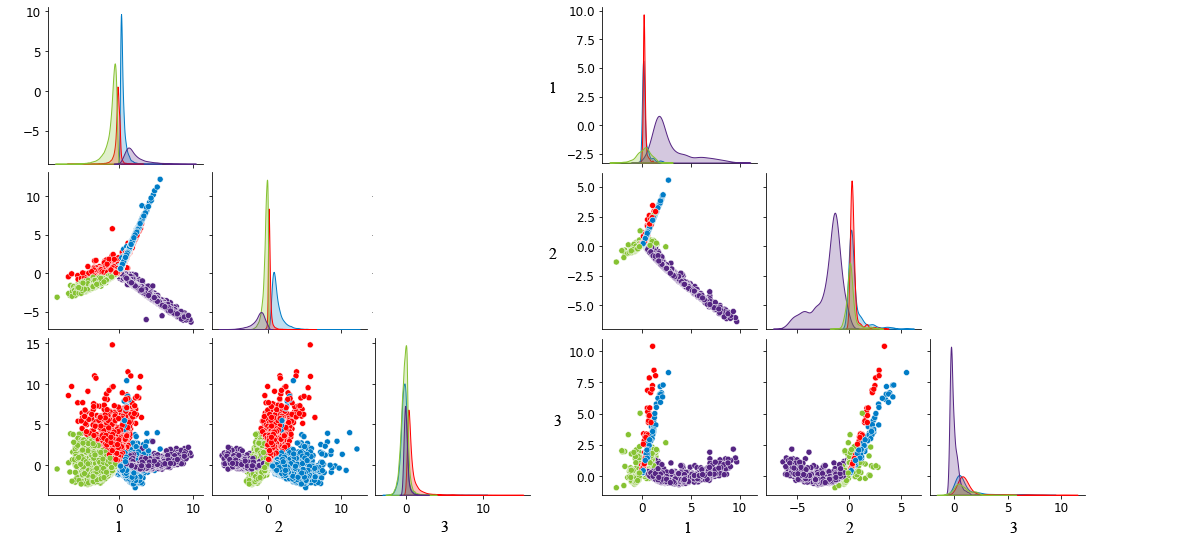}
    \caption{User clusters based on retweeting behavior within the latent sharing space. Left-hand plots include all users; right-hand plots are coordination candidates with $\phi \geq 0.67$. Cluster memberships are from Figure \ref{fig:size_clust}. Density plots for each are on the respective diagonals.}
    \label{fig:lat_clust}
\end{figure*}

The distribution of users within the latent space of the user-retweet matrix is displayed in Figure \ref{fig:lat_clust}. A scree plot from the SVD indicated three main dimensions of variation. User scores were clustered on these using HDBSCAN, resulting in a four-cluster solution; more dimensions input into HDBSCAN tended to break up the four clusters into a much larger number that were not interpretively useful. 

\begin{figure}
    \centering
    \includegraphics[width=0.8\textwidth]{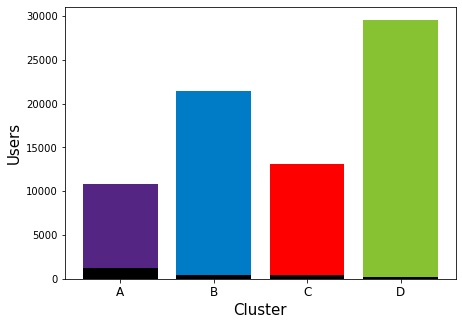}
    \caption{Bar plot of cluster sizes. Shaded areas correspond to the number of coordinated users found in each respective cluster.}
    \label{fig:size_clust}
\end{figure}

User scores for the first three dimensions are plotted pairwise in Figure \ref{fig:lat_clust}, colored according to the four-cluster solution; the scores for all users appear on the left, while those for only the coordination candidates are on the right. The three plots along the diagonal edge indicate the densities of the different clusters on each of the dimensions. Figure \ref{fig:size_clust} provides a key to cluster size; the labels and colors there are used for reference throughout. At the bottom of each bar, a black bar segment indicates the number of users with links above the 0.66 association threshold, i.e., those we consider candidates for coordinated activity. Not shown here, but relevant to fully appreciating the space in terms of user preferences for content and influencers to retweet, are the tweet loadings in the latent sharing space. 


The pair-wise plot in the top left corner compares user scores on dimensions 1 ($x$-axis) and 2 ($y$-axis). Dimension 1 most strongly separates clusters A and B (right) from C and D (left); dimension 2 primarily separates A from B, leaving C and D near the origin. This is also seen in the second row right-most plot, with dimension 2 ($x$-axis) and 3 ($y$-axis); C and D become more separated on dimension 3, with some contribution from dimension 2; a dip in density not fully aligned with the axes separates these groups. 

The clusters are mapped onto the user positions in the 3-nearest neighbor network layout in Figure \ref{fig:phi_netclust}: A turns out to be the highly-connected small cluster nearest the top; B is the larger cluster toward the right and the largest cluster is composed of C and D overlain on each other, with the smaller cluster D more dispersed around C. There is a scattering of users from all four clusters throughout the layout; several small satellite clusters are assigned to C in the latent space clustering, as well as a fringe of users close to B. Hence, density clustering identifies largely the same clustering as the Gephi layout, which otherwise employs information from dimensions not used in the latent space clustering. Hence, the latent sharing space of retweeting identifies different clusters of users with different connectivity properties, as previously noted. These are explored in turn in subsections below. 

\subsubsection{\color{violet} Cluster A}

\begin{figure}
    \centering
    \includegraphics[width=1\textwidth]{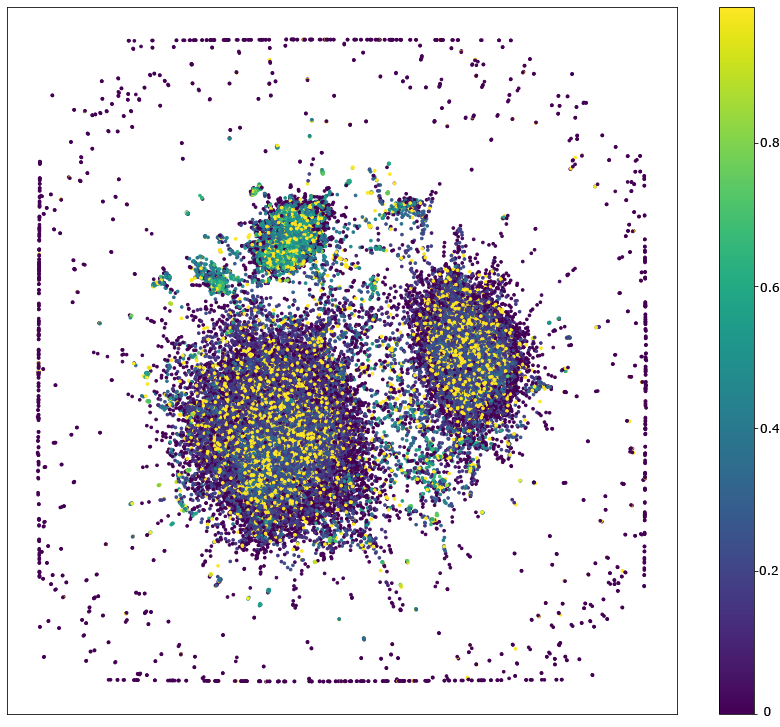}
    \caption{K-nearest neighbor network (k=3) where edges are weighted by $\phi$ and nodes are colored according by their largest edge weight. Coordinating nodes are distributed throughout the core and periphery of the K-NN graph. See \ref{fig:phi_netd} to compare with node degrees in Gram matrix of Matrix A.}
    \label{fig:phi_net}
\end{figure}

\begin{figure}
    \centering
    \includegraphics[width=1\textwidth]{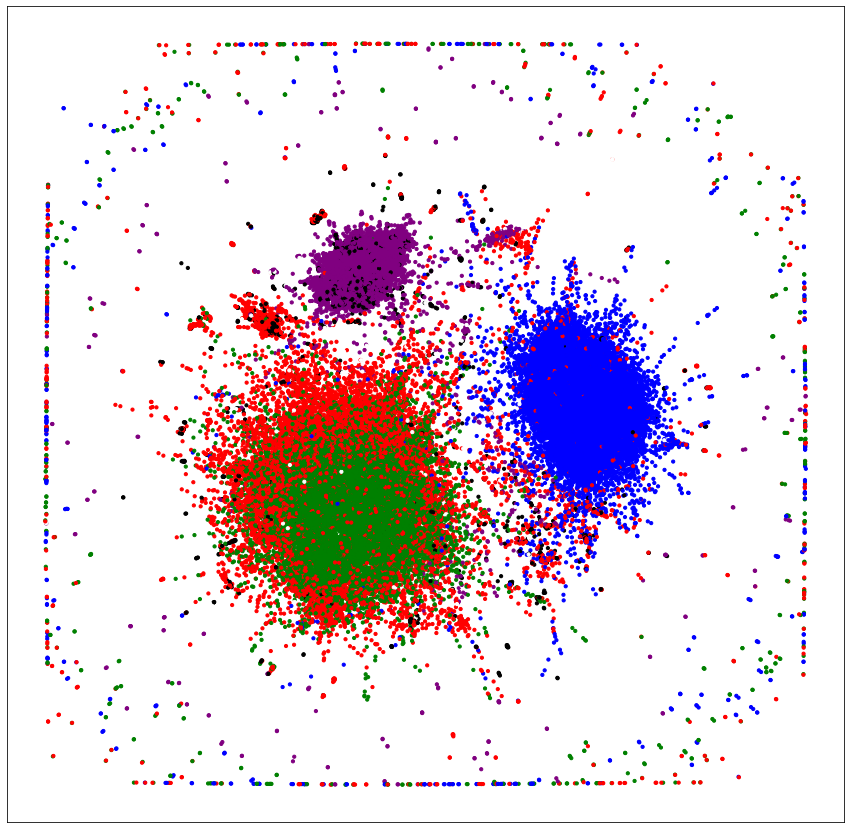}
    \caption{K-nearest neighbor network (k=3) where edges are weighted by $\phi$ and nodes are colored by their cluster membership, except for coordinated users who are shaded black for contrast.}
    \label{fig:phi_netclust}
\end{figure}

This cluster consists of 9,652 users engaged in activity promoting the South Korean band Bangtan Boys (BTS) and to a lesser extent other related bands such as Blackpink. The midterm election period in 2022 was in temporal proximity to the Peoples Choice Awards (PCA) and the American Music Awards (AMA). A very small number of users in this cluster shared content from Joe Biden, Barack Obama, and OccupyDemocrats in addition to their K-pop promotion. The users in this cluster engage in non-covert calls to action for promoting BTS, including explicit calls to retweets posts and post formulaic replies conveying the intent to and voting for the band in the People's Choice Awards and the American Music Awards. Sometimes a motivating user requests a certain number of retweets and replies from other users engaging in this promotional activity. In addition, the People's Choice Awards and American Music Awards have a voting system within Twitter, explaining the distinctive connectivity for these users as seen in Figure \ref{fig:phi_net} and Figure \ref{fig:phi_netclust}. 

Cluster A contains  the largest number of coordinated user candidates in our data, with 1,212 users accounting for more than one in ten users in this cluster. Posts shared by these candidates engage in the same band promotional and voting activity. Although these accounts do not engage in fundamentally different activity from non-candidate users in Cluster A, the degree of association between these users indicate higher levels of cooperation among this subset than is typical for music award voters. As such, this cluster is a good example of how self-organized behavior by users can occur alongside coordinated actors attempting to game some aspect of the organic activity.

\subsubsection{\color{Blue} Cluster B}
This cluster is composed of 20,965 users on the US political right, prominently retweeting political figures involved in election discourse. The most prominently shared post in this cluster is a post implying systematic voter fraud by the Democratic Party: " 52\% of Fetterman's votes were mail in ballots. Do you understand what they do now?" This form also holds for a number of other prominently shared posts including the second most prominent post: "Over 500,000 of Kathy Hochuls votes were mail in ballots pre-filled, all you had to do is sign. Do you understand what they do now?"

Cluster B contains 424 coordination candidates advocating for the Republicans and against Democrats. The most prominently shared user by these users, accounting for the first and second most shared posts, is currently suspended on Twitter/X. The third most shared post discusses feeling déjà vu with respect to the 2022 and 2020 election cycles and fraudulent activity. Other posts build on this theme by claiming that lines and other forms of chaos at voting stations are an intentional strategy to push people to vote by mail, with the implication that this will be another vector for voter fraud. Another prominent theme is the purported impending communist takeover of the United States and, in a similar vein, references to the arrival of a metaphorical storm or apocalyptical battle as articulated among Q-Anon members. 

\subsubsection{\color{BrickRed} Cluster C}
This cluster consists of 12,764 users on the political left who share a variety of Democrat-aligned messages, including from users such as POTUS (Biden administration) and Barack Obama. The most frequently shared posts in this cluster focus on abortion rights, showcasing examples of women who have historically voted Republican but plan to vote Democrat due to abortion legislation. Another prominently shard post is authored by Independent candidate Everett Stern imploring his would be supporters to vote, instead, for John Fetterman.

Cluster C contains 381 coordination candidates. Although Cluster C is largely political in nature, much of the coordinated activity is focused on promoting  and voting for Taylor Swift in the American Music Awards, much like the K-pop promotion and voting in Cluster A. Also in parallel to coordination candidates in Cluster A, there are explicit calls to mobilize retweets and replies: "I'm voting for Red (Taylor's Version) for Favorite Pop Album at the \#AMAs GOAL: 2k replies + 1k rts[.]" In addition, there is a small number of posts about Taylor Swift's support for the Democrats and abortion. Notably, users promoting Taylor Swift refer to "Red" and urge people to "Vote Red" in reference to one of her albums. Without additional treatment, a text-based analysis or clustering could be confounded by the divergent use of these references by Swift and Republican supporters, respectively.

\subsubsection{\color{Green} Cluster D}
This cluster consists of 29,278 users of a similar political left-leaning alignment to users in Cluster C, sharing some of the same influencers, including the aforementioned post by Everett Stern. It deviates from Cluster C with rhetoric that is more partisan and mocking of Republicans and personalities on the right. Cluster C, in contrast, has a greater tendency to share content from political analysts and commentators who refer to election related statistics and post breaking results from the midterms. 

Cluster D has 258 coordination candidates engaged in mobilizing support and voter turnout for Democratic candidates. One of the prominent themes promoted by these users include the role of millennials and Gen Z in preventing a ''red wave.'' This includes lobbying these demographic groups and providing explanations for why they should, and later explaining why they did, turn out for Democrats. Issues highlighted included gun violence and school shootings, abortion rights, and their dislike for Lauren Boebert. Hillary Clinton and Barack Obama are prominently retweeted by these users, with the former focusing on the threat to the United State's political system: "They're looking to impose one-party rule with zero accountability to voters, and they're not being subtle about it." Obama, on the other hand, is invoked for voter mobilization because "In many places, your vote could make the difference."  

\section{Discussion}

As in other contentious and important events over the last decade, the 2022 midterm elections were an arena where various kinds of influence campaigns were expressed in social media, including on Twitter \cite{pierriUkr,cov_twitter_v_facebook,Pacheco_2020}. Our analysis was driven by similar questions and challenges with respect to identifying these actions. Following widely-used approaches, we analyzed a set of Twitter data for signatures of coordinated activity in the retweeting network. Our approach employed a latent sharing space of retweeting behaviors, and a 3-nearest neighbor network of association strengths among users to identify coordination candidates and situate them within the larger context of communication. 

Association was found to be bimodal among the 3-nearest neighbor links of each user. Additionally, different levels of association corresponded to different kinds of connectivity structure within and between clusters in the network, allowing us to take a more informed approach to thresholding for identifying coordination candidates. In our analysis of communication patterns, three dimensions of latent sharing space were found in which four clusters of users were identified, having two distinct kinds of retweeting activity: political campaigning and music awards. Cluster A features a large number of users organizing around promoting K-pop bands for music awards. This seemingly innocuous activity also features the largest number of coordination candidates who, on the surface, appear to accomplish much of the same work as organic users in this cluster. This underscores the complexity of separating coordinated users from non-coordinated users who might share a common motivation but not a latent organizational relationship as is the case between coordinated users.  

Although our study makes an effort to understand coordinated activity alongside the larger retweet communication environment of the midterm elections, there are several limitations to our approach. By design, we scope our analysis to retweet activity, though coordination may have occurred along any number of additional behavioral dimensions. In addition, queries like the one used here \cite{aiyappa2023multi}, are inherently limited by the inevitability of capturing some content unrelated to the intended subject of inquiry, alongside the inability to capture absolutely all content relevant for an analysis. Usually, further treatment of what is captured is needed to properly account for the substance and motivations of the communication as those processes all bear on the structure of the interaction networks observed in the data. 

The presence of coordinated actions or campaigns on social media remains a matter of interest to researchers of the digital public square, not least when these organized actors attempt to sow divisions in our societies and undermine our collective faith in institutions we depend on for the basic functioning of our polity. Analyses of coordinated activity can also help avoid mistaking benign activities for nefarious ones, as well as avoid hazards for scientific inference. Researchers using social media data for social scientific studies or research on emergent social phenomenon may benefit from properly accounting for non-organic structures in the data. A clear example of this hazard in our study is Taylor Swift promoters and voters (Cluster C) who write "vote red" in reference to her album, readily confused with the same language used by Republicans (Cluster B). Failing to exclude or otherwise account for such different types of activity and motivations may bias our understanding of the data and measurement of self-organizing phenomena. In addition, narrow attention to the network structure without appreciating the distinct forms of activity within it may also lead to unsafe inferences about the nature and purpose of coordinated activity.

\begin{credits}
\subsubsection{\ackname} This study was supported in part by a grant from the Knight Foundation to the Indiana University Observatory on Social Media (OSoMe).
This work was also enabled by Jetstream2 at Indiana University from the Advanced Cyberinfrastructure Coordination Ecosystem: Services and Support (ACCESS) program, which is supported by National Science Foundation grants \#2138259, \#2138286, \#2138307, \#2137603, and \#2138296.
\end{credits}
%
%
%
 \bibliographystyle{splncs04}
 \bibliography{bib}

\begin{thebibliography}{10}
\providecommand{\url}[1]{\texttt{#1}}
\providecommand{\urlprefix}{URL }
\providecommand{\doi}[1]{https://doi.org/#1}

\bibitem{aiyappa2023multi}
Aiyappa, R., DeVerna, M.R., Pote, M., Truong, B.T., Zhao, W., Axelrod, D., Pessianzadeh, A., Kachwala, Z., Kim, M., Seckin, O.C., et~al.: A multi-platform collection of social media posts about the 2022 us midterm elections. In: Proceedings of the international AAAI conference on web and social media. vol.~17, pp. 981--989 (2023)

\bibitem{barbera2015birds}
Barber{\'a}, P.: Birds of the same feather tweet together: Bayesian ideal point estimation using twitter data. Political analysis  \textbf{23}(1),  76--91 (2015)

\bibitem{Barbera_2}
Barberá, P., Jost, J., Nagler, J., Tucker, J., Bonneau, R.: Tweeting from left to right: Is online political communication more than an echo chamber? Psychological Science  \textbf{26},  1531--1542 (2015)

\bibitem{ICWSM09154}
Bastian, M., Heymann, S., Jacomy, M.: Gephi: An open source software for exploring and manipulating networks. In: International AAAI Conference on Weblogs and Social Media (2009), \url{http://www.aaai.org/ocs/index.php/ICWSM/09/paper/view/154}

\bibitem{bergsma2013bias}
Bergsma, W.: A bias-correction for cram{\'e}r’s v and tschuprow’s t. Journal of the Korean Statistical Society  \textbf{42}(3),  323--328 (2013)

\bibitem{cima2024coordinated}
Cima, L., Mannocci, L., Avvenuti, M., Tesconi, M., Cresci, S.: {Coordinated behavior in information operations on Twitter}. IEEE Access  (2024)

\bibitem{conover2011predicting}
Conover, M.D., Gon{\c{c}}alves, B., Ratkiewicz, J., Flammini, A., Menczer, F.: Predicting the political alignment of twitter users. In: 2011 IEEE third international conference on privacy, security, risk and trust and 2011 IEEE third international conference on social computing. pp. 192--199. IEEE (2011)

\bibitem{tuckerRussia}
Eady, G., Paskhalis, T., Zilinsky, J., Bonneau, R., Nagler, J., Tucker, J.: Exposure to the russian internet research agency foreign influence campaign on twitter in the 2016 us election and its relationship to attitudes and voting behavior. Nature Communications  \textbf{14}(62) (2023)

\bibitem{PolPolar2023}
Flamino, J., Galeazzi, A., Feldman, S., Macy, M., Cross, B., Zhou, Z., Serafino, M., Bovet, A., Makse, H., Szymanski, B.: Political polarization of news media and influencers on twitter in the 2016 and 2020 us presidential elections. Nature Human Behaviour  \textbf{7},  904–916 (2023)

\bibitem{coURL}
Gabriel, N.A., Broniatowski, D.A., Johnson, N.F.: Inductive detection of influence operations via graph learning (2023)

\bibitem{magelinski2022synchronized}
Magelinski, T., Ng, L., Carley, K.: A synchronized action framework for detection of coordination on social media. Journal of Online Trust and Safety  \textbf{1}(2) (2022)

\bibitem{mcinnes2017accelerated}
McInnes, L., Healy, J.: Accelerated hierarchical density based clustering. In: Data Mining Workshops (ICDMW), 2017 IEEE International Conference on. pp. 33--42. IEEE (2017)

\bibitem{nizzoli2021coordinated}
Nizzoli, L., Tardelli, S., Avvenuti, M., Cresci, S., Tesconi, M.: Coordinated behavior on social media in 2019 uk general election. In: Proceedings of the International AAAI Conference on Web and Social Media (2021)

\bibitem{nwala2023language}
Nwala, A.C., Flammini, A., Menczer, F.: {A Language Framework for Modeling Social Media Account Behavior}. EPJ Data Science  \textbf{12}(1), ~33 (2023)

\bibitem{Pacheco_2020}
Pacheco, D., Flammini, A., Menczer, F.: Unveiling coordinated groups behind white helmets disinformation. In: Companion Proceedings of the Web Conference 2020 (2020). \doi{10.1145/3366424.3385775}, \url{https://doi.org/10.1145/3366424.3385775}

\bibitem{Pacheco_2021}
Pacheco, D., Hui, P.M., Torres-Lugo, C., Truong, B.T., Flammini, A., Menczer, F.: Uncovering coordinated networks on social media: Methods and case studies. Proc. International AAAI Conference on Web and Social Media  \textbf{15}(1),  455--466 (2021)

\bibitem{pierriUkr}
Pierri, F., Luceri, L., Jindal, N., Ferrara, E.: Propaganda and misinformation on facebook and twitter during the russian invasion of ukraine. In: Proceedings of the 15th ACM Web Science Conference 2023. p. 65–74. WebSci '23, Association for Computing Machinery, New York, NY, USA (2023), \url{https://doi.org/10.1145/3578503.3583597}

\bibitem{saeed2024unraveling}
Saeed, M.H., Ali, S., Paudel, P., Blackburn, J., Stringhini, G.: {Unraveling the Web of Disinformation: Exploring the Larger Context of State-Sponsored Influence Campaigns on Twitter}. arXiv preprint arXiv:2407.18098  (2024)

\bibitem{mutlifacet2020}
Tardelli, S., Nizzoli, L., Avvenuti, M., Cresci, S., Tesconi, M.: Multifaceted online coordinated behavior in the 2020 us presidential election. EPJ Data science  \textbf{13} (2024)

\bibitem{tardelli2023temporal}
Tardelli, S., Nizzoli, L., Tesconi, M., Conti, M., Nakov, P., Martino, G.D.S., Cresci, S.: Temporal dynamics of coordinated online behavior: Stability, archetypes, and influence. arXiv preprint arXiv:2301.06774  (2023)

\bibitem{wong2016quantifying}
Wong, F.M.F., Tan, C.W., Sen, S., Chiang, M.: Quantifying political leaning from tweets, retweets, and retweeters. IEEE transactions on knowledge and data engineering  \textbf{28}(8),  2158--2172 (2016)

\bibitem{cov_twitter_v_facebook}
Yang, K.C., Pierri, F., Hui, P.M., Axelrod, D., Torres-Lugo, C., Bryden, J., Menczer, F.: The covid-19 infodemic: Twitter versus facebook. Big Data \& Society  \textbf{8}(1),  20539517211013861 (2021). \doi{10.1177/20539517211013861}, \url{https://doi.org/10.1177/20539517211013861}

\bibitem{zannettou2019disinformation}
Zannettou, S., Caulfield, T., De~Cristofaro, E., Sirivianos, M., Stringhini, G., Blackburn, J.: {Disinformation Warfare: Understanding State-sponsored Trolls on Twitter and their Influence on the Web}. In: Companion Proc. WWW Conf. pp. 218--226 (2019). \doi{10.1145/3308560.3316495}

\end{thebibliography}
\end{document}